# Local and Global Dynamics of a Functionally Graded Dielectric Elastomer Plate

**Amin Alibakhshi**[1], **Sasan Rahmanian**[2], **Michel Destrade**[3,4,], **Giuseppe Zurlo**[4]

[1] Escuela Técnica Superior de Ingeniería Aeronáutica y del Espacio, Universidad Politécnica de Madrid, Pza. Cardenal Cisneros 3, 28040, Madrid, Spain

Email address: amin.alibakhshi@upm.es

[2] Department of Systems Design Engineering, University of Waterloo, Waterloo, N2L 3G1, Canada

Email address: s223rahm@uwaterloo.ca

[3] Key Laboratory of Soft Machines and Smart Devices of Zhejiang Province & Department of Engineering Mechanics & Soft Matter Research Center, Hangzhou 310027, P.R. China

[4] School of Mathematical and Statistical Sciences, University of Galway, University Road, Galway, Republic of Ireland

Email address: michel.destrade@universityofgalway.ie

Email address: giuseppe.zurlo@universityofgalway.ie

**Abstract** We investigate the nonlinear vibrations of a functionally graded dielectric elastomer plate subjected to electromechanical loads. We focus on local and global dynamics in the system. We employ the Gent strain energy function to model the dielectric elastomer. The functionally graded parameters are the shear modulus, mass density, and permittivity of the elastomer, which are formulated by a common through-thickness power-law scheme. We derive the equation of motion using the Euler-Lagrange equations and solve it numerically with the Runge-Kutta method and a continuation-based method. We investigate the influence of the functionally graded parameters on equilibrium points, natural frequencies, and static/dynamic instability. We also establish a Hamiltonian energy method to detect safe regions of operating gradient parameters. Furthermore, we explore the effect of the functionally graded parameters on chaos and resonance by plotting several numerical diagrams, including time histories, phase portraits, Poincaré maps, largest Lyapunov exponent criteria, bifurcation diagram of Poincaré maps, and frequency-stretch curves. The results provide a benchmark for developing functionally graded soft smart materials.

## 1. Introduction

Polymers (elastomers and rubbers) are some of the most widely used soft materials in many systems. They display material nonlinearity (nonlinear strain-stress curve) and can sustain large deformations. Smart materials often rely on the so-called active polymers (Dorfmann & Ogden, 2014; Kim & Tadokoro, 2007; Liu et al., 2013; Meng & Hu, 2010; Xia et al., 2021), especially dielectric elastomers (DEs) (Behera et al., 2021; Guo et al., 2021; Gupta & Harursampath, 2015; Jiang et al., 2021; Khurana et al., 2021, 2022).

DEs are active polymers that deform nonlinearly in response to electromechanical loads (Lu et al., 2020; Suo, 2010; Zhao et al., 2011), and in some applications where external excitations are time-dependent, they may generate nonlinear oscillations (Wang et al., 2016). DEs are designed as soft and flexible membranes with diverse geometries (beams (Alibakhshi, Dastjerdi, Fantuzzi, 2022; Feng et al., 2011), square and rectangular (Conroy Broderick et al., 2020; Xia et al., 2021; Zhang et al., 2018), tubular and cylindrical shells (Bazaev & Cohen, 2022; Bortot & Shmuel, 2018; Ghosh & Basu, 2021; Ren & Guo, 2021; Su, 2020), spherical shells (Lv et al., 2018; Yong et al., 2011)), with surfaces covered by compliant and flexible electrodes subjected to static or/and dynamic voltages through the thickness (Tommasi et al., 2014). In response, DEs expand in the in-plane direction and shrink in the thickness direction. Sometimes, to enhance their performance, a mechanical load (equal-biaxial, uniaxial, pure shear) is also applied (Zhang & Chen, 2020). The resulting nonlinear vibrations can be complex and accompanied by chaos, quasiperiodicity, and instability, depending on the operating parameters.

For instance, Zhu *et al.* (Zhu et al., 2010) investigated the nonlinear vibration of a spherical DE membrane modelled by the neo-Hookean hyperelastic model with deformation-independent permittivity, and derived the governing equations by the virtual work method. Xu *et al.* (Xu et al., 2012) analyzed the nonlinear dynamics of a thick-walled square DE neo-Hookean membrane with the Euler-Lagrange equations. Sheng *et al.* (Sheng et al., 2014) investigated the nonlinear vibrational response of a thin-walled DE membrane with strain-stiffening and damping effects. Alibakhshi and co-workers (Alibakhshi, Chen, et al., 2022; Alibakhshi, Dastjerdi, Akgöz, et al., 2022; Alibakhshi, Dastjerdi, Malikan, et al., 2022; Alibakhshi et al., 2021) studied the nonlinear vibration of DEs with different geometries and different hyperelastic models, including the Gent, neo-Hookean, Gent-Gent, and generalized neo-Hookean models. Cooley and Lowe (Cooley & Lowe, 2022) studied the nonlinear resonance of a circular DE using Hamilton's principle, and analyzed the natural frequency and frequency amplitude response. Zou *et al*. (Zou et al., 2022) presented a dynamic analysis of circular DEs with a focus on chaotic oscillations; they used the Gent strain energy function for modeling the nonlinearity and strain-stiffening effect, and employed the Lyapunov exponent to identify chaotic domains.

Hence, there is a large volume of works investigating the nonlinear vibrations and dynamics of homogeneous DEs. However, it might be beneficial to consider that DEs could be designed as functionally graded materials, to improve on their performance. So far, only a limited number of works have treated the stability and modeling of functionally graded DEs (FGDEs). Su *et al*. (Su et al., 2021) investigated the bending of an FGDE plate caused by voltage, assuming that the elastic shear modulus and the electric permittivity vary linearly within the thickness. Zhou *et al*. (Zhou et al., 2020) assessed the bifurcation response of a FGDE tube under axial stretch deformation and radial electric potential, also assuming a linear variation of the parameters in the thickness direction. Chen and Yang (Chen & Yang, 2021) analyzed the performance of a FGDE energy harvester disc with graded material parameters in the radial direction. Alam and Sharma (Alam & Sharma, 2022) studied longitudinal wave band gaps in a FGDE modelled by a compressible neo-Hookean strain energy function and power law functionally graded rule; they used a finite element approach in conjunction with Bloch-Floquet theory. Wu *et al*. (Wu et al., 2020) investigated the propagation of axisymmetric waves in a tube modelled as a (linearly) functionally graded Mooney-Rivlin ideal dielectric.

Here we conduct a nonlinear vibration analysis on such FGDE structures and as a test case, we focus on the membrane geometry. We use the Gent hyperelastic model with a power law for the functional gradient in the thickness direction, common to all material parameters. In Section 2 we present this material and derive the governing equation of motion using the Euler-Lagrange equations. In Section 3 we conduct a detailed analysis of the free and forced vibration regimes, with special attention to local and global dynamics. The outcomes of our analysis disclose new possibilities to broaden the working range of dielectric elastomers, above all for those applications that are crucially based on the dynamical behavior of these devices (for example, energy harvesting (Colonnelli et al., 2015; De Tommasi et al., 2014; Tommasi et al., 2014; Zurlo et al., 2018)).

## 2. Mathematical modelling

Fig. 1 is the schematic representation of a functionally graded dielectric elastomer (FGDE) planar membrane. The upper and lower surfaces are coated with compliant electrodes, which are subjected to an electric potential difference $V$. The membrane is also under an in-plane equi-biaxial tensile mechanical load $P$. It undergoes large deformations, from the undeformed, unloaded reference configuration to the current configuration where the membrane is deformed under external electromechanical loadings. The associated Cartesian coordinate systems are $(X_1, X_2, X_3)$ and $(x_1, x_2, x_3)$, respectively, and the length, width, and thickness are $L$, $L$, $H$, and $l$, $l$, $h$, respectively.

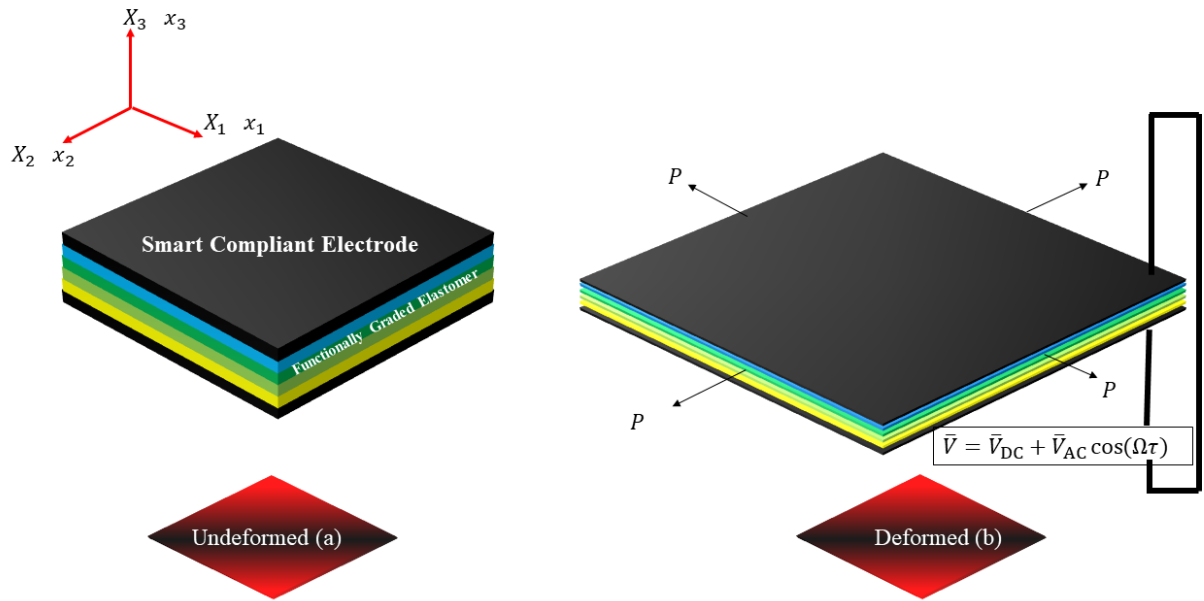

**Figure. 1** Schematic representation of a functionally graded dielectric elastomer. The elastomer is very thin and the effect of inertia in the thickness direction is neglected. In sub-figures, we amplified the thickness for representation purpose only.

The principal stretches are defined as

$$\lambda_1(t) = \frac{x_1}{X_1}, \qquad \lambda_2(t) = \frac{x_2}{X_2}, \qquad \lambda_3(t) = \frac{x_3}{X_3} \tag{1}$$

in the $x_1$, $x_2$, and $x_3$ directions, respectively. We assume an in-plane equi-biaxial deformation, so that $\lambda_1 = \lambda_2 = \lambda$, and incompressibility, so that $\lambda_1 \lambda_2 \lambda_3 = 1$. It follows that

$$\lambda(t) = \frac{x_1}{X_1}, \qquad \lambda(t) = \frac{x_2}{X_2}, \qquad \lambda^{-2} = \frac{x_3}{X_3} \tag{2}$$

Next we define the kinetic energy of the system as (Xu et al., 2012)

$$\mathcal{T} = \int_0^{x_3} \int_0^{x_2} \int_0^{x_1} \frac{1}{2} \rho \left( \dot{x_1}^2 + \dot{x_2}^2 + \dot{x_3}^2 \right) dx_1 dx_2 dx_3 \tag{3}$$

where the dot denotes differentiation with respect to time. We assume that the mass density changes smoothly and continuously in the thickness direction, according to the following power FG law (Pascon, 2018)

$$\rho(X_3) = \rho_2 + (\rho_1 - \rho_2) \left( \frac{X_3}{H} \right)^K \tag{4}$$

where $\rho_1$ is the density at $X_3 = H$ and $\rho_2$ is the density at $X_3 = 0$, and $K$ is the gradient index. Then we get

$$\mathcal{T} = \int_0^H \int_0^L \int_0^L \frac{1}{2} \left( \rho_2 + (\rho_1 - \rho_2) \left( \frac{X_3}{H} \right)^K \right) \left( \left( X_1 \dot{\lambda} \right)^2 + \left( X_2 \dot{\lambda} \right)^2 \right) \underbrace{\lambda_1 \lambda_2 \lambda_3}_{1} \, dX_1 dX_2 dX_3 \tag{5}$$

after changing the coordinate variables from spatial to material. The result of the integration is

$$\mathcal{T} = \frac{1}{3} \frac{(\rho_1 + K\rho_2)}{(1 + K)} HL^4 \dot{\lambda}^2 \tag{6}$$

where we neglected terms of orders higher than $(H/L)^3$, and also neglected the inertia in the thickness direction because the membrane is thin (effectively taking $\dot{x_3} = 0$).

The total potential energy is

$$\mathcal{U} = \int_0^{x_3} \int_0^{x_2} \int_0^{x_3} (\psi_H + \psi_E) \, dx_1 dx_2 dx_3 - W_P \tag{7}$$

where $\psi_H$ is the strain energy density, $\psi_E$ is the electric field potential and $W_P$ is the work done by the lateral tensile mechanical load.

Dielectric elastomers materials are commonly described as having rubber-like and elastomeric properties, as they can undergo large deformations and display a nonlinear relationship between stress and strain. To model this behavior, several hyperelastic constitutive laws have been introduced, with the neo-Hookean and Gent models being the most used for describing the elastic properties of dielectric elastomers. The neo-Hookean model is appropriate for small-to-moderate deformations, and the Gent model captures the strain-stiffening phenomenon observed at large stretches, for example by commercially available dielectric elastomers such as 3M's VHB 4905 and VHB 4910. According to the Gent model (Gent, 1996),

$$\psi_H = -\frac{\mu J_m}{2} \ln\left(1 - \frac{I_1 - 3}{J_m}\right) = -\frac{\mu J_m}{2} \ln\left(1 - \frac{2\lambda^2 + \lambda^{-4} - 3}{J_m}\right) \tag{8}$$

where $\mu$ is the infinitesimal shear modulus and $I_1 = \lambda_1^2 + \lambda_2^2 + \lambda_3^2$ is the first invariant of deformation. The dimensionless strain-stiffening parameter $J_m$ gives a measure of the limiting stretch $\lambda_{\text{lim}}$, which is found as the real root of $2\lambda^2 + \lambda^{-4} - 3 = J_m$. We point out that other, similar, hyperelastic models such as the Gent-Gent model (Mangan & Destrade, 2015), the generalized neo-Hookean model (Anssari-Benam & Bucchi, 2021; Horgan, 2021), the Lopez-Pamies model (Zurlo et al., 2018), etc., can be used for modelling the nonlinear elastic response of DEs.

We assume that $\mu$ is a functionally graded parameter, following the same power law as the density (Pascon, 2018)

$$\mu(X_3) = \mu_2 + (\mu_1 - \mu_2)\left(\frac{X_3}{H}\right)^K \tag{9}$$

where $\mu_1$, $\mu_2$ are the shear moduli at $X_3 = H$, $X_3 = 0$, respectively. Substituting Eq. (9) into Eq. (8), and integrating with a change of variables, we obtain the total strain energy $W_H$ as

$$W_H = \int_0^H \int_0^L \int_0^L \psi_H \underbrace{\lambda_1 \lambda_2 \lambda_3}_{1} \, dX_1 dX_2 dX_3 = -\frac{(\mu_1 + K\mu_2)J_m H L^2}{2(1+K)} \ln\left(1 - \frac{2\lambda^2 + \lambda^{-4} - 3}{J_m}\right) \tag{10}$$

Now we write the electric part of the potential energy as

$$\psi_E = -\frac{1}{2}\varepsilon\left(\frac{V}{H}\right)^2 \lambda^4 \tag{11}$$

where $\varepsilon$ denotes the permittivity of the DE. Again, we assume a power-law variation across the thickness, as

$$\varepsilon(X_3) = \varepsilon_2 + (\varepsilon_1 - \varepsilon_2)\left(\frac{X_3}{H}\right)^K \tag{12}$$

say. Using Eqs. (11) and (12), the total electric potential energy is obtained by integration over the volume of the membrane and change of variables, as

$$W_E = -\frac{(\varepsilon_1 + \varepsilon_2 K)H L^2}{2(1+K)}\left(\frac{\phi}{H}\right)^2 \lambda^4 \tag{13}$$

We also compute the work done by the external tensile load, as

$$W_P = \int_{X_1}^{x_1} P \, dx_1 + \int_{X_2}^{x_2} P \, dx_2 = P(x_1 - X_1) + P(x_2 - X_2) = 2PL(\lambda - 1) = 2\sigma L^2 H(\lambda - 1) \tag{14}$$

in which $\sigma = P/LH$ stands for the stress caused by the mechanical load.

Finally, putting together these contributions, we find that the total potential energy of the system is

$$\mathcal{U} = -\frac{(\mu_1 + \mu_2 K)J_m HL^2}{2(1+K)} \ln\left(1 - \frac{2\lambda^2 + \lambda^{-4} - 3}{J_m}\right) - \frac{(\varepsilon_1 + \varepsilon_2 K)HL^2}{2(1+K)}\left(\frac{V}{H}\right)^2 \lambda^4$$
$$- 2\sigma HL^2(\lambda - 1) \tag{15}$$

We may now derive the equation of motion, using the Euler-Lagrange equation (other methods include the virtual work principle and Hamilton's principle (Yin et al., 2022)). Hence we write that (Amabili, 2008)

$$\frac{d}{dt}\left(\frac{\partial \mathcal{L}}{\partial \dot{\lambda}}\right) - \frac{\partial \mathcal{L}}{\partial \lambda} = 0 \tag{16}$$

where $\mathcal{L} = \mathcal{T} - \mathcal{U}$ is the Lagrangian. Here, we arrive at

$$\frac{L^2 \rho_1}{3\mu_1}(1 + nK)\frac{d^2\lambda}{dt^2} + (1 + rK)\frac{J(\lambda - \lambda^{-5})}{J - 2\lambda^2 - \lambda^{-4} + 3}$$
$$- \frac{\varepsilon_1(1 + Km)}{\mu_1}\left(\frac{V}{H}\right)^2 \lambda^3 - \frac{(K+1)\sigma}{\mu_1} = 0 \tag{17}$$

where $n = \rho_2/\rho_1$, $r = \mu_2/\mu_1$, and $m = \varepsilon_2/\varepsilon_1$ are the inhomogeneity ratios. We introduce the following non-dimensional measures of time, voltage and mechanical load,

$$\tau = \frac{t}{L\sqrt{\rho_1/3\mu_1}}, \qquad \bar{V} = \sqrt{\varepsilon_1/\mu_1}\left(\frac{V}{H}\right), \qquad \bar{P} = \frac{\sigma}{\mu_1} \tag{18}$$

so that the non-dimensional version of the equation is

$$(1 + nK)\frac{d^2\lambda}{d\tau^2} + (1 + rK)\frac{J_m(\lambda - \lambda^{-5})}{(J_m - 2\lambda^2 - \lambda^{-4} + 3)} - (1 + mK)\bar{V}^2\lambda^3 - (1 + K)\bar{P} = 0 \tag{19}$$

consistent with the homogeneous DE (Sheng et al., 2014) case when $K = 0$.

## 3. Static and DC dynamic responses

In this section we study the static response of the membrane to applied quasistatic mechanical loads and static voltages. Then we investigate how it behaves dynamically when it is first pre-stretched by a quasistatic load followed a sudden (time-step) applied static voltage.

An important assumption here is that we restrict our attention to homogeneous deformations, both in the static and in the dynamic cases. Due to the non-monotonicity of the voltage-stretch curves, this assumption means that we do not consider the possible occurrence of phase transitions in the membrane, where thin and thick regions ("phases") may coexist inside the spinodal region of the loading curve. The assumption that the electromechanically actuated membrane deforms homogeneously simplifies the study, as the system is then described by a single degree-of-freedom variable (the planar stretch). The richer scenario with phase transitions, both in the static and in the dynamic regimes, is remarkably more complex and we leave this task for a future study.

Following an energy-based method proposed by Sharma *et al* (Sharma et al., 2018), we first write $\bar{U} = \mathcal{U}/L^2 H \mu_1$, the nondimensional form of the total potential energy, as

$$\bar{U} = -\frac{(1+rK)J_m}{2(1+K)} \ln\left(1 - \frac{2\lambda^2 + \lambda^{-4} - 3}{J_m}\right) - \frac{(1+rK)}{2(1+K)}\bar{V}^2\lambda^4 - 2\bar{P}(\lambda - 1) \qquad (20)$$

We then write the balance of static equilibrium as

$$\frac{d\bar{U}}{d\lambda} = (1+rK)\frac{J_m(\lambda - \lambda^{-5})}{(J_m - 2\lambda^2 - \lambda^{-4} + 3)} - (1+mK)\bar{V}^2\lambda^3 - (1+K)\bar{P} = 0 \qquad (21)$$

and find critical states of instability/stability by solving

$$\frac{d^2\bar{U}}{d\lambda^2} = (1+rK)J_m\frac{(1+5\lambda^{-6})}{(J-2\lambda^2-\lambda^{-4}+3)} + (1+rK)J_m\frac{4(\lambda-\lambda^{-5})^2}{(J-2\lambda^2-\lambda^{-4}+3)^2} - 3(1+mK)\bar{V}^2\lambda^2 = 0. \qquad (22)$$

By solving Eqs. (21) and (22) simultaneously, we obtain two critical values $\lambda_C^S$ and $\bar{V}_C^S$. Accordingly, we obtain the static instability actuation stretch $\lambda_{ac}^S = \lambda_C^S/\lambda_P$ where $\lambda_P$ is the stretch when there is no voltage applied, obtained by solving

$$(1+rK)\frac{J_m(\lambda_P - \lambda_P^{-5})}{(J_m - 2\lambda_P^2 - \lambda_P^{-4} + 3)} - (1+K)\bar{P} = 0 \qquad (23)$$

To plot the $\bar{V} - \lambda$ curves, we solve numerically the algebraic Eq. (21), see examples in Fig. 2. The Figure and Tables 1, 2, 3 display values of the critical voltage corresponding to the local maximum on the loading curve, which we refer to as the "limit point" voltage, in reference to the terminology employed for the inflation of rubber balloons. For the DE plates, it corresponds to the vanishing of the Hessian of the free energy (Norris, 2008), and also to the onset of thin-plate inhomogeneous instability (Su et al., 2018).

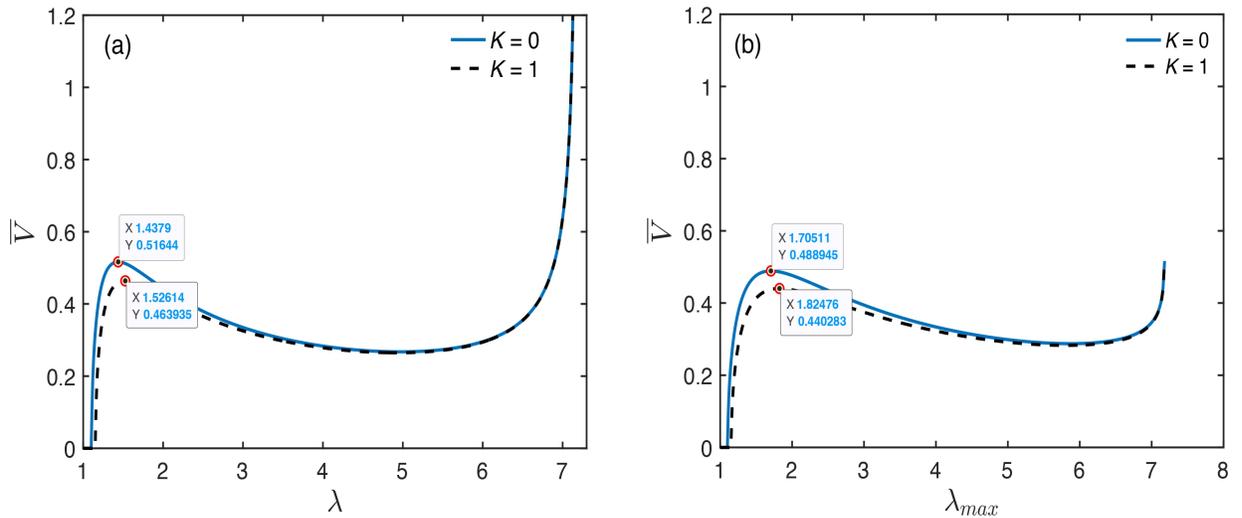

Now we incorporate the effect of inertia and motion into the system, by considering that a quasistatic mechanical pre-stretch is applied, followed by a DC Heaviside step voltage, so that the system responds with dynamic vibrations (Sharma et al., 2017). The ordinary differential equation governing these motions is Eq. (19).

Again, we use an energy-based approach: we formulate the (non-dimensional) Hamiltonian of the system as $\bar{\mathcal{H}} = \bar{\mathcal{U}} + \bar{\mathcal{T}}$ (where $\bar{\mathcal{T}} = \frac{\mathcal{T}}{3\mu_1 L^2 H}$ is the dimensionless measure of the kinetic energy) as

$$\bar{\mathcal{H}}(\tau) = \frac{(1+Kn)}{(1+K)}\left(\frac{d\lambda}{d\tau}\right)^2 - \frac{(1+rK)J_m}{2(1+K)}\ln\left(1 - \frac{2\lambda^2 + \lambda^{-4} - 3}{J_m}\right) - \frac{(1+mK)}{2(1+K)}\bar{V}^2\lambda^4 - 2\bar{P}(\lambda - 1) \tag{24}$$

Because the system is conservative, the Hamiltonian is constant. Its value at time $\tau = 0$ (DE at rest, $\lambda(0) = \lambda_P, \dot{\lambda}(0) = 0$) and at time $\tau = \tau_2$ (DE at maximum overshoot, $\lambda(\tau_2) = \lambda_{max}, \dot{\lambda}(\tau_2) = 0$) is the same (Sharma et al., 2017), so that

$$\begin{aligned}
\bar{\mathcal{D}}(\tau_2) = \bar{\mathcal{H}}(\tau_2) - \bar{\mathcal{H}}(0) \\
= -\frac{(1+rK)J_m}{2(1+K)}\ln\left(\frac{J_m + 3 - 2\lambda_{max}^2 - \lambda_{max}^{-4}}{J_m + 3 - 2\lambda_P^2 + \lambda_P^{-4}}\right) - \frac{(1+mK)}{2(1+K)}\bar{V}^2(\lambda_{max}^4 - \lambda_P^4) \\
- 2\bar{P}(\lambda_{max} - \lambda_P) = 0
\end{aligned} \tag{25}$$

This equation can be used to plot the $\lambda_{max} - \bar{V}$ curves of dynamic loading, see Fig.2b.

To find the critical DC dynamic overshoot stretch $\lambda_C^D$ and corresponding voltage $V_C^D$, we solve $\partial\bar{\mathcal{D}}/\partial\lambda_{max} = 0$, which gives

$$(1+rK)\frac{J_m(\lambda_{max} - \lambda_{max}^{-5})}{(J_m - 2\lambda_{max}^2 - \lambda_{max}^{-4} + 3)} - (1+mK)\bar{V}^2\lambda_{max}^3 - (1+K)\bar{P} = 0 \tag{26}$$

Simultaneously solving Eq. (25) and (26) gives $\lambda_{max} = \lambda_C^D$ and $V_C^D$, and consequently, the actuation stretch in the dynamic mode as $\lambda_{ac}^D = \lambda_C^D/\lambda_P$.

In Table 1, we present the effect of the gradient parameter $K$ on the static and DC dynamic limit point voltage. We increase $K$ as $K = 0.0, 0.25, 1.0, 2.5$. In the static case, we see that $\lambda_P$ increases, the critical voltage $V_C^S$ decreases, and the critical stretch $\lambda_C^S$ increases. In the dynamic case, the trend is similar: the critical dynamic stretch $\lambda_C^D$ increases while the dynamic critical voltage $\bar{V}_C^D$ decreases.

**Table. 1** Effect of parameter $K$ on the limit point voltage (static and dynamic cases), with $n = r = m = 0.5$.

| | Static | | Dynamic | |
|---|---|---|---|---|
| **$K = 0$** | $\lambda_P = 1.10538$ | | | |
| | $\bar{V}_C^S = 0.51644$ | $\lambda_C^S = 1.4379$ | $\bar{V}_C^D = 0.488945$ | $\lambda_C^D = 1.70511$ |
| **$K = 0.25$** | $\lambda_P = 1.12062$ | | | |
| | $\bar{V}_C^S = 0.49853$ | $\lambda_C^S = 1.46503$ | $\bar{V}_C^D = 0.472343$ | $\lambda_C^D = 1.74176$ |
| **$K = 1$** | $\lambda_P = 1.15393$ | | | |
| | $\bar{V}_C^S = 0.463935$ | $\lambda_C^S = 1.52614$ | $\bar{V}_C^D = 0.440283$ | $\lambda_C^D = 1.82476$ |
| **$K = 2.5$** | $\lambda_P = 1.19153$ | | | |
| | $\bar{V}_C^S = 0.431275$ | $\lambda_C^S = 1.59797$ | $\bar{V}_C^D = 0.41002$ | $\lambda_C^D = 1.92278$ |

In Table 2 we look at the influence of the shear modulus gradient parameter $r = \mu_2/\mu_1$, and in Table 3, at the influence of the permittivity gradient parameter $m = \varepsilon_2/\varepsilon_1$. An important conclusion gleaned from the tables is that the dynamic critical limit point stretch is greater than the static critical voltage, allowing for a greater expansion of the plate before the onset of instability.

**Table. 2** Effect of parameter $r$ on the limit point voltage (static and dynamic cases), with $n = m = 0.5$, and $K = 1$.

| | Static | | Dynamic | |
|---|---|---|---|---|
| **$r = 0.3$** | $\lambda_P = 1.18847$ | | | |
| | $\bar{V}_C^S = 0.403763$ | $\lambda_C^S = 1.59202$ | $\bar{V}_C^D = 0.383809$ | $\lambda_C^D = 1.91465$ |
| **$r = 0.5$** | $\lambda_P = 1.15393$ | | | |
| | $\bar{V}_C^S = 0.463935$ | $\lambda_C^S = 1.52614$ | $\bar{V}_C^D = 0.440283$ | $\lambda_C^D = 1.82476$ |
| **$r = 0.8$** | $\lambda_P = 1.12062$ | | | |
| | $\bar{V}_C^S = 0.546112$ | $\lambda_C^S = 1.46503$ | $\bar{V}_C^D = 0.517426$ | $\lambda_C^D = 1.74176$ |

**Table. 3** Effect of parameter $m$ on the limit point voltage (static and dynamic cases), with $n = r = 0.5$ and $K = 1$.

| | Static instability | | Dynamic instability | |
|---|---|---|---|---|
| **$m = 0.3$** | $\lambda_P = 1.15393$ | | | |
| | $\bar{V}_C^S = 0.498346$ | $\lambda_C^S = 1.52614$ | $\bar{V}_C^D = 0.47294$ | $\lambda_C^D = 1.82476$ |
| **$m = 0.5$** | $\lambda_P = 1.15393$ | | | |
| | $\bar{V}_C^S = 0.463935$ | $\lambda_C^S = 1.52614$ | $\bar{V}_C^D = 0.440283$ | $\lambda_C^D = 1.82476$ |
| **$m = 0.8$** | $\lambda_P = 1.15393$ | | | |
| | $\bar{V}_C^S = 0.423513$ | $\lambda_C^S = 1.52614$ | $\bar{V}_C^D = 0.401922$ | $\lambda_C^D = 1.82476$ |

In Fig. 3, we study the DC dynamic response numerically by plotting the $\lambda - \tau$ (time history) and $\mathrm{d}\lambda/\mathrm{d}\tau - \lambda$ (phase plane) curves. Here, we implement a Runge-Kutta method for the numerical time-integration of Eq. (19) with the initial conditions that the membrane is pre-stretched and at rest when the DC voltage is applied, i.e. $\lambda(0) = \lambda_P$ and $\dot{\lambda}(0) = 0$.

Fig.3 shows what happens when the applied voltage approaches the limit point voltage, in the cases $n = r = m = 0.5$, and $K = 0, 1$. When it is just below the limit point value identified in the tables ($\bar{V}_C^D = 0.440283$), the motion (full lines) is periodic and smooth, and centred around the static stretch corresponding to lowest root of Eq. (21) ($\bar{V}_C^S = 0.463935$). When it is just above, the membrane stretches extensively, to reach the maximum overshoot on the stiffening branch of the dynamic loading $\lambda_{max} - \bar{V}$ curve, given by Eq.(25). It then oscillates periodically between that extreme value and $\lambda_P$, undergoing large amplitude motions (shown in Fig. 3 using dashed lines). Fig. 3 also shows the corresponding phase planes.

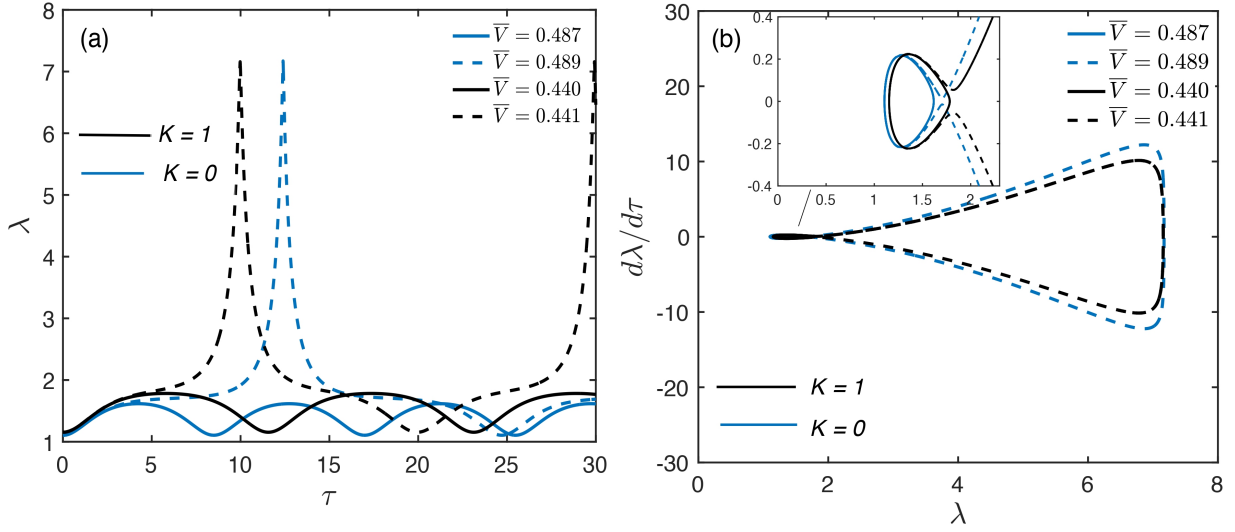

**Figure. 3** Effect of functionally graded parameters on the dynamic motion of the FGDE with $J_m = 100$ and $\bar{P} = 0.5$. **(a-b)** Time history and phase plane diagram for $n = r = m = 0.5$, and $K = 0, 1$. Full/dashed lines: the DC voltage is just below/above the dynamic limit point voltage.

## 4. Nonlinear resonance analysis

In this section we investigate the resonances caused by external and parametric excitations for the fixed parameters $n = r = m = 0.5$ and $J_m = 100$.

To find the natural frequencies of the system, we re-write the equation of motion Eq. (19) in normal form, as $d^2\lambda/d\tau^2 + \mathcal{M}(\lambda(\tau)) = 0$, where

$$\mathcal{M}(\lambda) = \frac{(1 + K\,r)}{(1 + K\,n)} \frac{J_m(\lambda - \lambda^{-5})}{(J_m - 2\lambda^2 - \lambda^{-4} + 3)} - \frac{(1 + Km)}{(1 + K\,n)} \bar{V}^2 \lambda^3 - \frac{(K + 1)}{(1 + K\,n)} \bar{P} \qquad (27)$$

The equilibrium stretch $\lambda_{eq}$ is found by solving $\mathcal{M}(\lambda_{eq}) = 0$, which is the same as solving Eq. (21). Following the approach of Zhu *et al.* (Zhu et al., 2010), we look for small-amplitude oscillations in the neighborhood of $\lambda_{eq}$, as $\lambda(\tau) = \lambda_{eq} + x(\tau)$, where $x(\tau)$ is small. Then the equation of motion, $d^2\lambda/d\tau^2 + \mathcal{M}(\lambda) = 0$, can be linearised to $d^2x/d\tau^2 + \omega_n^2 x = 0$, where

$$\omega_n = \sqrt{\frac{\partial \mathcal{M}}{\partial \lambda}(\lambda_{eq})} \qquad (28)$$

is the dimensionless *natural frequency*.

We first focus on the role played by the tensile prestress $P$ on the natural frequency, as illustrated in Fig. 4 for different values of the applied static voltage $\bar{V}$. When the plate is homogeneous ($K=0$), and the voltage is zero, the natural frequency attains its minimum value $\omega_n \approx 1.143$ when $\bar{P} \approx 2.055$.

Application of non-zero voltage acts both qualitatively and quantitatively on the dynamic response of the dielectric membrane. For increasing values of the applied voltage, the U shape of the curve is substantially preserved, but the minimum value of $\omega_n$ decreases. Eventually, the voltage reaches a critical value $\bar{V} \approx 0.2$, as the prestress increases to a maximum value $P \approx 1.9$. From Eqs. (27) and (28), and comparing with Eqs. (21) and (22), we see that solving $\omega_n = 0$ is equivalent to finding the critical limit point voltage. From the dynamical perspective, reaching the limit point voltage means that the condition $\omega_n^2 \geq 0$ is no longer satisfied and $\omega_n$ becomes a purely imaginary quantity. Then vibrations with amplitude proportional to $e^{\pm i\omega_n t}$ may grow unbounded, and the small-amplitude assumption is no longer valid. In that sense, the limit point voltage corresponds to an instability, at least in the linearized sense. Fig. 4 also shows the influence of the functional gradient, when $K = 1.0$, $n = m = r = 0.5$.

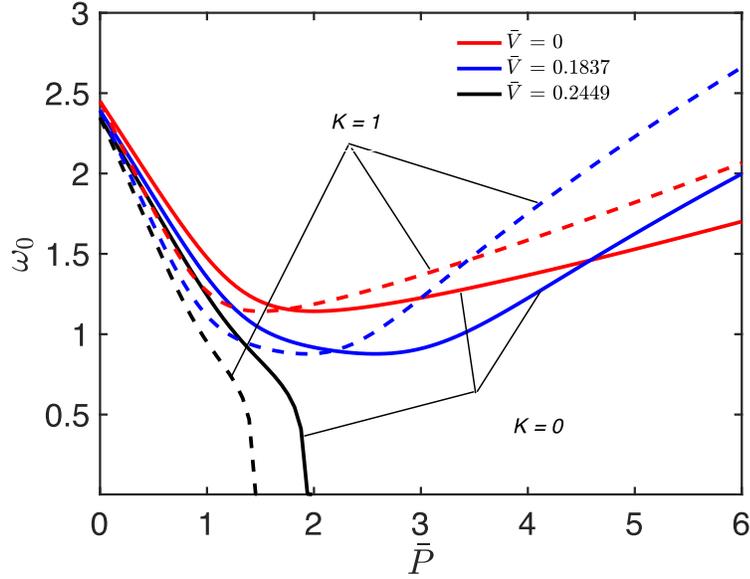

**Figure 4**. Variations of the natural frequency of small-amplitude oscillations of the dielectric plate with the pre-stress load, for different values of the static voltage $\bar{V} = 0, 0.18, 0.25$ when the plate is homogeneous ($K = 0$), and when it is functionally graded ($K = 1.0$, $n = m = r = 0.5$).

The frequency-response behavior of the membrane in the proximity of the primary resonance is analysed in Fig. 5, by solving the nonlinear equation of motion The frequency-displacement curve begins with small-amplitude oscillations, smoothly followed by a small peak near half the natural frequency, corresponding to "superharmonic resonance". As the excitation frequency approaches the natural frequency, the stable small-amplitude solution branch undergoes a cyclic-fold (CF) bifurcation, thus merging with the unstable solution branch at $\Omega \approx 2.19$, which is marked by $\mathtt{CF_1}$. Following a backward sweep at this point, the system response proceeds along the unstable periodic orbit, where the oscillation amplitude increases as the excitation frequency decreases, until yet another cyclic-fold bifurcation point is reached (not shown in the figure). This happens because the length of the stable solution branch that merges with the unstable branch is extremely short.

To capture the right stable solution branch, the excitation frequency is set initially to 2.8 and is then reduced to 2.2 by backward sweep. We thus see that at high frequencies and far away from the resonance region, the response amplitude is attracted by small-amplitude periodic orbits, whose amplitude grows gradually as the excitation frequency is decreased. Following this branch, the system dynamics loses its stability at a frequency close to 2.2, which is denoted by CF$_2$. Note that the two bifurcation points CF$_1$ and CF$_2$ shown in the figure are not exactly on top of each other, meaning that there is a narrow frequency bandwidth in which there is no stable periodic response.

The impact of pre-stress on the steady-state behaviour of the FGDE is illustrated in Fig. 5a. This figure shows that a small tensile pre-stress causes the bifurcation points to shift towards lower frequencies, an effect that is enhanced at higher frequencies. The impact of the power-law index for the functionally graded material is illustrated in Fig. 5b. Oppositely to the case where prestress is controlled, we now see that an increase in $K$ results into shifting the bifurcation points toward higher frequencies; furthermore, the shift is frequency-independent, meaning that the bandwidth remains now constant.

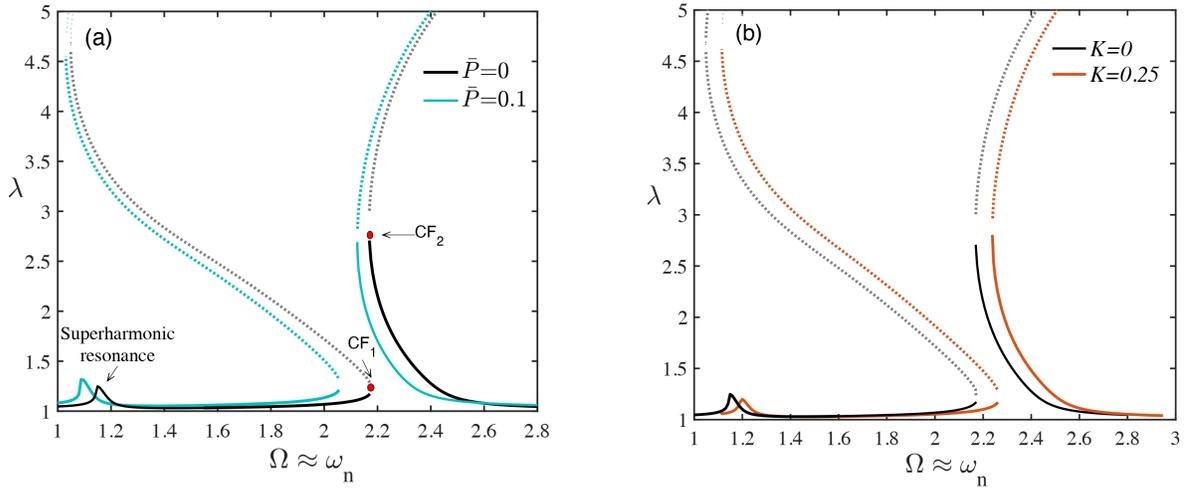

**Figure 5.** Frequency-response behavior of the FGDE near its primary resonance for $\bar{V}_{DC} = \bar{V}_{AC} = 0.2$, for (a) two different values of pre-stress load, (b) two different values of power law index.

We now focus on the parametric resonance characteristics of the FGDE structure. To this end, the forcing frequency is set to be varying near twice the natural frequency of the structure, thus exciting the principal parametric resonance. Since here the forcing is proportional to the square of the applied voltage, (which is composed of both DC and AC components), two harmonic excitation terms are included in the system dynamics, one at the AC frequency ($\Omega$) itself, and one at two times the AC frequency ($2\Omega$). It is assumed that the AC frequency is swept near twice the AC frequency, making the first excitation term to excite the principal parametric resonance, whereas the second harmonic term does not produce resonance. For this reason, the activation of the principal parametric resonance is caused only by the first harmonic excitation term, whose frequency is the same as the AC signal frequency.

Fig. 6 illustrates the steady-state dynamics of the FGDE structure in the neighbour of its principal parametric resonance. As shown in Fig. 6a, in the absence of pre-stress the frequency-displacement curve contains two almost horizontal solution branches, where frequency has only very mild variations. It is worth mentioning that the proposed system is under external excitation as well, creating an offset for the small-amplitude motion with respect to the non-zero static equilibrium. Compared to dynamic systems with equilibrium around zero, the horizontal branches correspond to the trivial solutions in the frequency-response curve.

We note that there is a bias component in the system response, originating from the non-zero equilibrium position. According to Fig. 6a, as the pre-stress is increased, the level of the horizontal solution branches rises, meaning that the position of the static equilibrium point is shifted towards higher values. It is also observed that both subcritical and the supercritical primary Hopf bifurcations (denoted by PH) occur. Another interesting observation is that the resonance frequency bandwidth becomes broader as the value of the pre-stress tensile force increases, meaning that the parametric resonance activation level drops.

On the other hand, the power-law index $K$ plays an opposite role on the parametric resonance characteristics. Fig. 6b shows the dependence of the parametric resonance activation level with the order of the functionally graded material. Now, as $K$ is increased, the bifurcation points loci shift towards higher frequencies, while the resonance bandwidth is reduced.

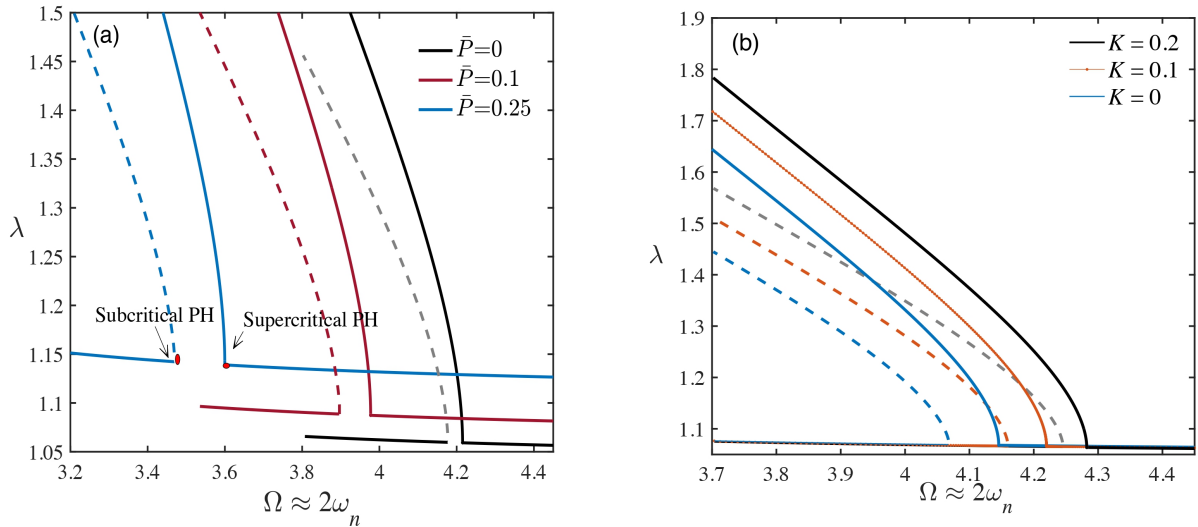

**Figure 6**. Frequency-response behavior of the FGDE near principal parametric resonance, for (a) three different values of pre-stress load with $\bar{V}_{DC} = \bar{V}_{AC} = 0.36$ and $K = 0$, (b) three different values of power law index with $\bar{V}_{DC} = \bar{V}_{AC} = 0.38$.

The time history of the system and its phase portrait are further analysed in Fig. 7a and 7b. To capture the occurrence of resonance, the excitation frequency is now set at twice the natural frequency, because in the application of a voltage in the form of DC+AC, the natural frequency changes because of the bias part of the applied voltage.

As is clear from Fig. 7, when the applied voltage is $\bar{V}_{DC} = \bar{V}_{AC} = 0.3$ (which is below the activation level of the parametric resonance), the response amplitude remains substantially close to the static equilibrium solution; however, when the voltage is further increased to 0.3+0.01, the response grows in amplitude until it reaches a stable large-amplitude orbit, with period equal to twice the forcing period.

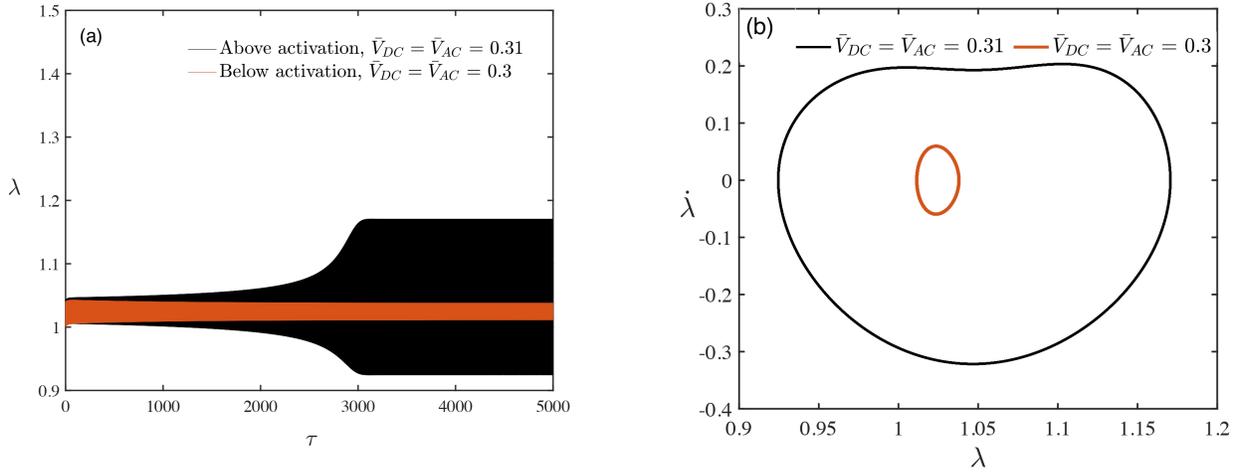

**Figure 7.** Dynamic response of the FGDE at $\Omega = 2\omega_0$, for two different values of the applied voltage, $\bar{V}_{DC} = \bar{V}_{AC} = 0.3$ and $\bar{V}_{DC} = \bar{V}_{AC} = 0.31$; (a) time history of the system response, (b) phase portrait of the steady-state response of the structure. In both cases, the gradient index is $K = 0$.

The frequency spectrum corresponding to both time-histories of Fig. 7 is analysed in Fig. 8 through a Fast Fourier Transform (FFT). For $\bar{V}_{DC} = \bar{V}_{AC} = 0.3$ (below the activation level of the parametric resonance) there is only one frequency peak at the forcing frequency ($\Omega \approx 2\omega_0$), corresponding to the main harmonic. When the parametric pump exceeds the activation level, $\bar{V}_{DC} = \bar{V}_{AC} = 0.3 + 0.01$, a second harmonic with large amplitude appears at $\frac{\Omega}{2} \approx \omega_0$, indicating that the response is now composed of two harmonics, with the dominant one taking place at half the forcing frequency. This is consistent with the results based on the inspection of the time history.

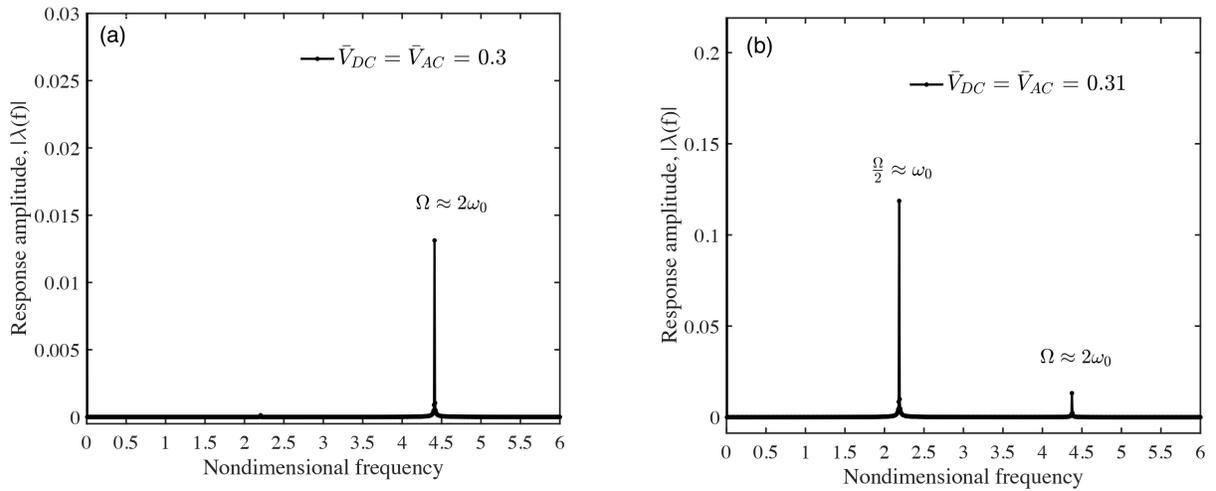

**Figure 8.** FFT plot of the dynamic resposne of the FGDE structure driven at $\Omega = 2\omega_0$, for two different values of the applied voltage; (a) $\bar{V}_{DC} = \bar{V}_{AC} = 0.3$, (b) $\bar{V}_{DC} = \bar{V}_{AC} = 0.31$. The gradient index is $K = 0$.

## 5. Chaotic oscillations

Due to the nonlinear nature of the behavior of DEs, chaotic oscillations can be expected (Zou et al., 2022). To investigate this possibility, we consider the following nondimensional applied voltage,

$$\bar{V} = \bar{V}_{DC} + \bar{V}_{AC}\cos(\Omega\tau) \tag{29}$$

where $\bar{V}_{DC}$ stands for the static voltage, $\bar{V}_{AC}$ denotes the amplitude of time-varying voltage, and $\Omega$ is the dimensionless excitation frequency (Zhu et al., 2010).

By substituting Eq. (29) into Eq. (19), we obtain the following ordinary differential equation for $\lambda$,

$$\frac{d^2\lambda}{d\tau^2} + \frac{(1+K\,r)}{(1+K\,n)}\frac{J_m(\lambda-\lambda^{-5})}{(J_m-2\lambda^2-\lambda^{-4}+3)} - \frac{(1+Km)}{(1+K\,n)}\bar{V}_{DC}^2\left(1+\frac{\bar{V}_{AC}}{\bar{V}_{DC}}\cos(\Omega\tau)\right)^2\lambda^3$$
$$-\frac{(K+1)}{(1+K\,n)}\bar{P} = 0 \tag{30}$$

After solving this equation through a Runge-Kutta method, the Poincaré maps (representing velocity $d\lambda/d\tau$ versus displacement $\lambda$) are obtained through sections of the response history, for time steps equal to $2\pi/\Omega$. This analysis is aimed at elucidating the role of the parameters $K$, $n$, $r$, $m$ on the onset of chaotic oscillations. For the other parameters we set $J_m = 100$, $\bar{P} = 0.5$, $\bar{V}_{DC} = \sqrt{0.1}$, $\bar{V}_{AC}/\bar{V}_{DC} = 0.35$. The results of the Poincaré maps are also paralleled by the study of the "Largest Lyapunov Exponent" (LLE) of the system – see (Amabili, 2018) for a thorough discussion on this method.

As can be appreciated from Table 1 to Table 3, the control parameters affect the equilibrium points and the natural frequency. Therefore, to study the response of the system we set $\Omega = 1.5$, which is far from the natural frequency of the system. Furthermore, we set $\lambda(\tau = 0) = 1.5$ for the initial stretch, which is far from the equilibrium state, and we prescribe null initial velocity.

We first focus on the role of the control parameter $K$ in the range between 0 and 2.55. For this analysis, the remaining inhomogeneity parameters are set as $n = r = m = 0.5$. The results, which are reported in Fig. 9a, highlight the presence of chaos for $K$ between 0 and 1.48. Past this range, the motion ceases to be chaotic and becomes quasiperiodic.

In the interval $0 < K < 1.48$ we obtain a cloud of points mapped on the Poincaré section, a sign of chaos. In contrast, for the quasiperiodic motion, the points appear as part of a closed curve. Another way of identifying the presence of chaos is to plot a vertical line in the bifurcation diagram for each value of the control parameter; if this line crosses a single point, this is a periodic motion; if it crosses in two points, we we have a periodic-2 motion.; etc. The chaotic behavior emerges if instead there is a large number of points on the vertical line.

Note that the outcomes of the bifurcation diagram are consistent with the analysis of the LLE, which is shown in Fig. 9b. We here recall that the value of the LLE is positive, zero and negative if the behavior of the system is chaotic, quasiperiodic and periodic, respectively.

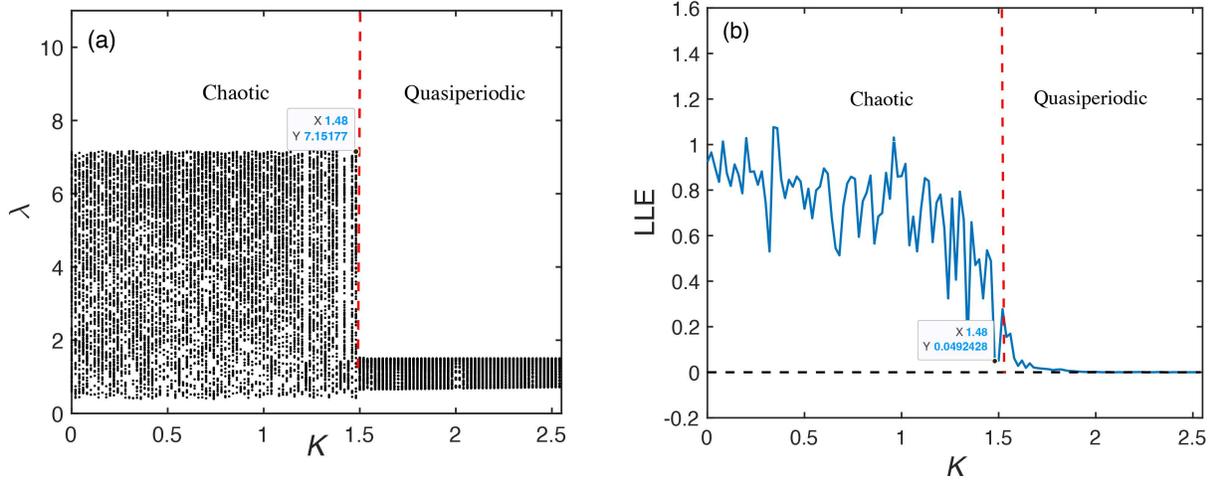

**Figure. 9** Nonlinear dynamic characteristics of the FGDE with $n = 0.5$, $r = 0.5$, and $m = 0.5$. (a) Bifurcation diagram of Poincaré map. (b) Largest Lyapunov exponent (LLE).

The transition between periodic, quasiperiodic and chaotic regimes can also be appreciated by studying the time history and the phase diagram. In Fig. 10a-b we report these plots for $K = 1$, whereas $K = 2.5$ is considered in Fig. 10c-d. These choices for $K$ are motivated by the conclusions of the frequency-response analysis (see Fig.6). When $K = 1$, the time history plotted in Fig. 10a has an irregular behavior due to chaos. This figure also shows the presence of electromechanical instabilities, such as snap-through and snap-back, which are evident due to abrupt jumps between between small and high stretches. These plots also highlight the bounding role of limit stretch extensibility of the Gent material. The Poincaré map for $K = 1$, which is obtained by sampling the excitation frequency $\Omega$ at every period (Ghayesh & Farokhi, 2015), is illustrated in Fig. 10b. An uncountable number points appear in this plot, again a signature of chaotic oscillations.

When the control parameter $K$ is increased to 2.5, the behavior becomes quasiperiodic, with the clear appearance of beatings. In the Poincaré map, the signature of quasiperiodicity is evident from the closedness of the response curve. We finally note, from Figs. 9 and 10, that an increase of $K$ corresponds to a reduction of the amplitude of oscillations.

Taken together, the conclusions of the analysis above show that the parameter $K$ plays a fundamental and well-defined role on the onset of chaos in the FGDE system.

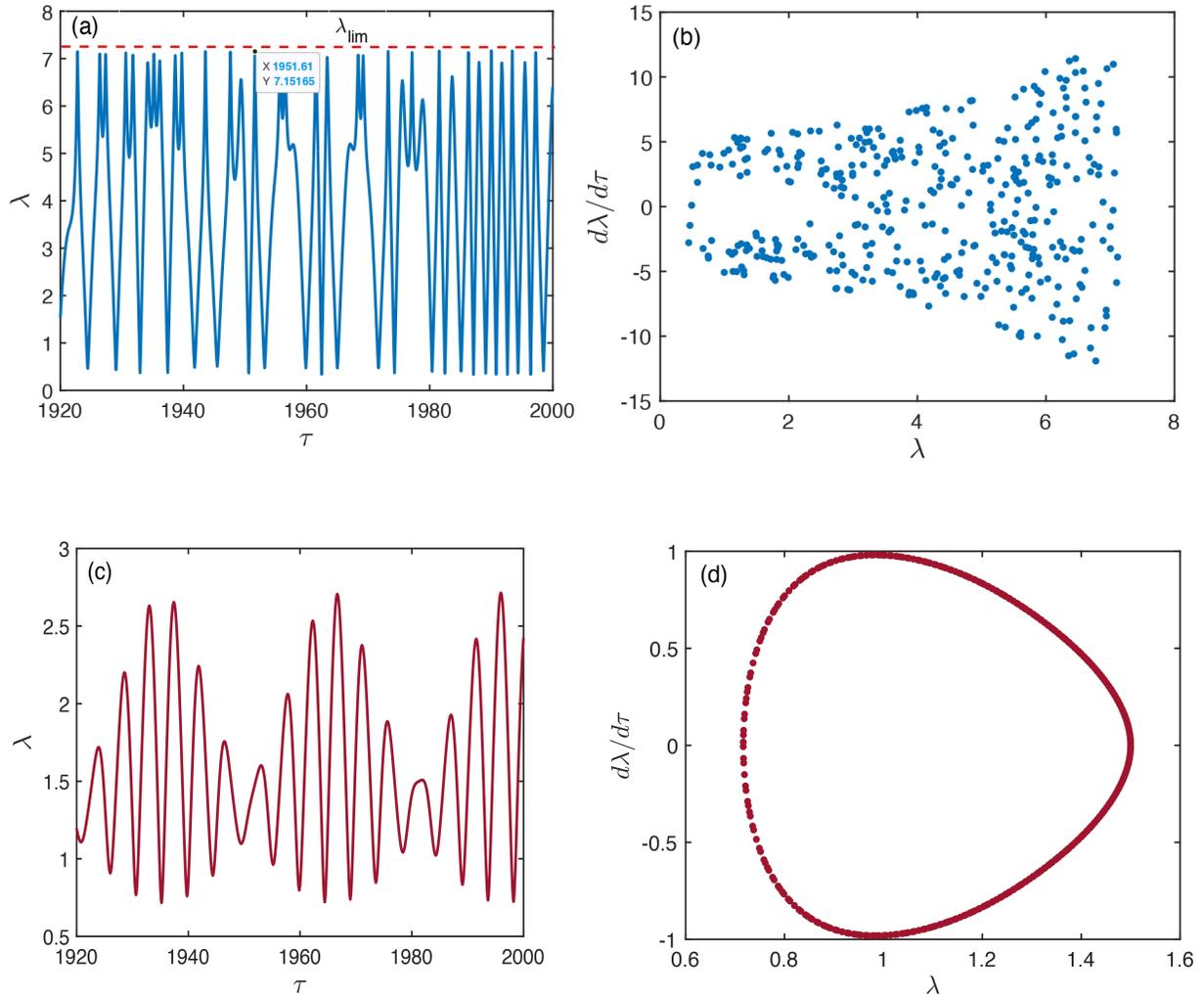

**Figure. 10** Nonlinear dynamic characteristics of the FGDE with $n = 0.5$, $r = 0.5$, and $m = 0.5$. (a-b) Time history and Poincaré map for $K = 1$. (c-d) Time history and Poincaré map for $K = 2.5$.

We now consider the influence of the control parameters $n$, $r$, $m$ on the dynamic behaviour of the system. Concerning the dependence on $n$, as illustrated from Fig. 11, we see that this parameter plays a role which is analogous to the role of $K$: oscillations are chaotic for $n$ between 0 and 0.54, and past this value the oscillations are quasiperiodic.

Far less trivial is the role played by the parameter $r$, as illustrated in Fig. 12. The domain for the controlling parameter $r$ is from 0 to 0.9. As $r$ increases, we find that windows of chaotic and quasiperiodic behaviour alternate.

Finally, the parameter $m$ does not seem to play a crucial role into the occurrence of chaos – at least, when the values of the remaining parameters are set in the range of interest in this manuscript. This feature can be appreciated by inspection of Fig. 13.

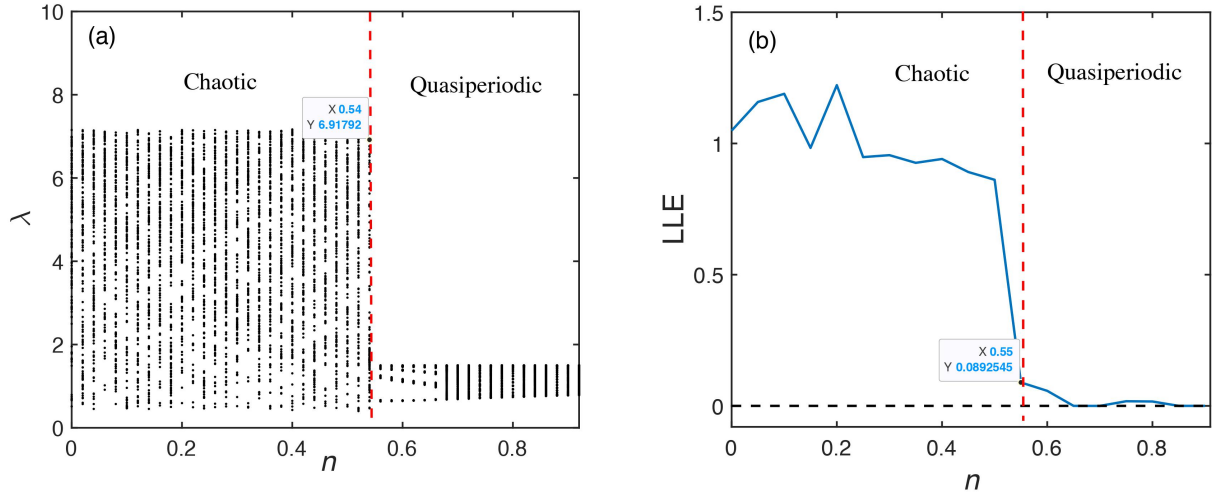

**Figure. 11** Nonlinear dynamic characteristics of the FGDE with $K = 1$, $r = 0.5$, and $m = 0.5$. (a) Bifurcation diagram of Poincaré map. (b) Largest Lyapunov exponent (LLE).

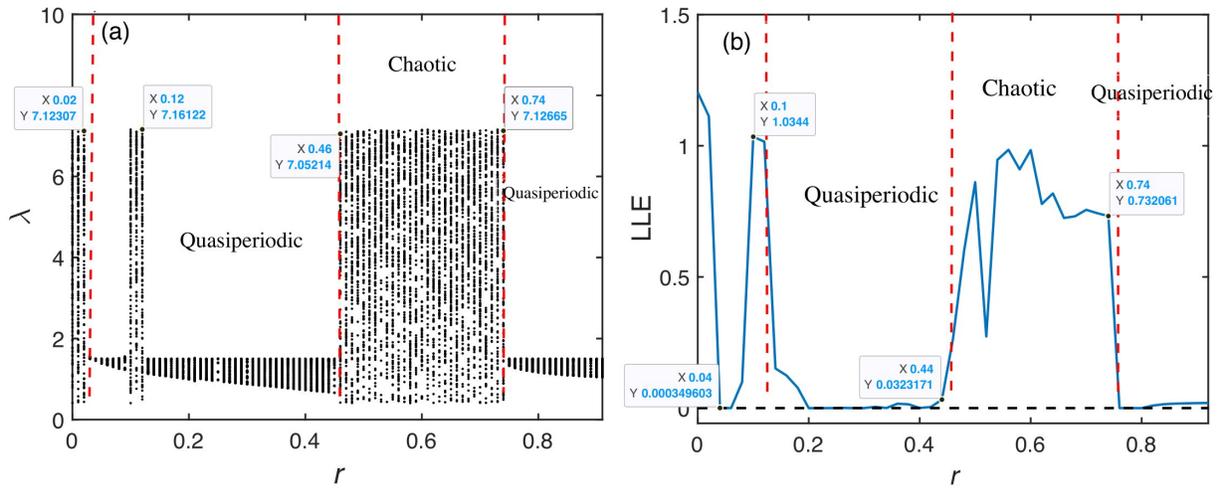

**Figure. 12** Nonlinear dynamic characteristics of the FGDE with $K = 1$, $n = 0.5$, and $m = 0.5$. (a) Bifurcation diagram of Poincaré map. (b) Largest Lyapunov exponent (LLE).

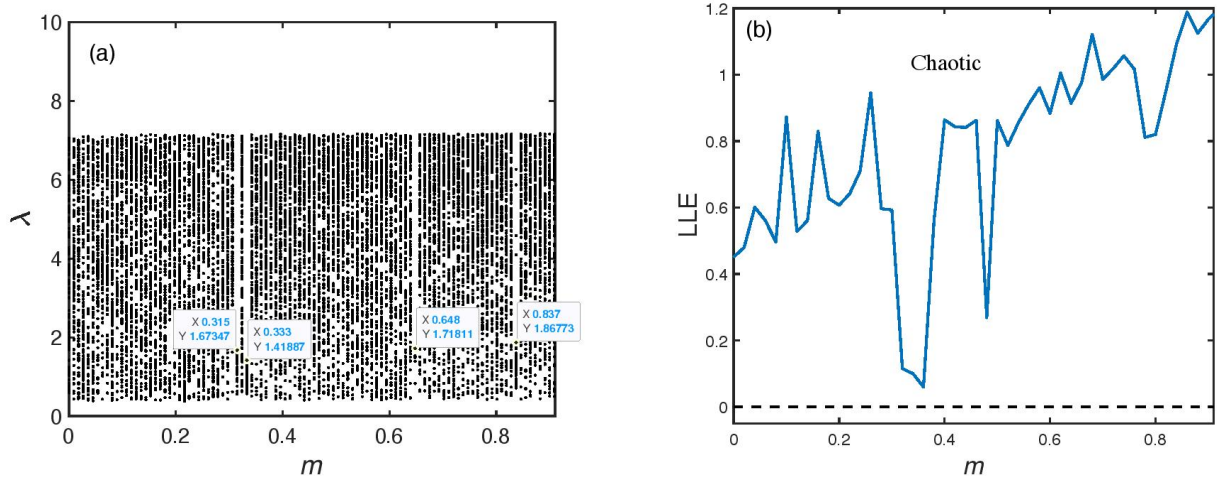

**Figure. 13** Nonlinear dynamic characteristics of the FGDE with $K = 1$, $n = 0.5$, and $r = 0.5$. (a) Bifurcation diagram of Poincaré map. (b) Largest Lyapunov exponent (LLE).

## 6. Conclusion

We studied the dynamical response of a functionally graded dielectric elastomer membrane, subjected to an in-plane tensile prestress and a time-varying voltage. The dielectric membrane is thin and incompressible, and its constitutive behavior is of the Gent type (strain-stiffening). The material is functionally graded in the thickness direction, with material parameters such as shear modulus, mass density and electrical permittivity varying according to a common power law.

We then studied the dynamic behavior of the system, by first deriving the equation of motion in terms of the in-plane stretch $\lambda(t)$, and then solving this equation numerically using a fourth-order Runge-Kutta method.

Our analysis is fundamentally aimed at elucidating the role played by the functionally graded parameters of the dielectric membrane on its static and dynamic behavior, with special emphasis on the occurrence of chaos. To this end, we produced voltage-stretch diagrams $\bar{V} - \lambda$, voltage-time diagrams $\lambda - \tau$, Poincaré maps, bifurcation diagrams of Poincaré maps, and Largest Lyapunov Exponent diagrams, and we have studied the resonance response of the system with the aid of shooting and arc-length continuation methods.

Our study ultimately leads to the following conclusions:

(1) Functional grading of the dielectric membrane deeply affects both its static and its dynamic behavior.
(2) Functional grading plays an important role on the snap-through and snap-back instabilities.
(3) Functional grading of the dielectric membrane has a very strong role on the emergence of chaos.
(4) Relative to the primary resonance, the system undergoes a cyclic fold bifurcation.
(5) Relative to the principal parametric resonance, both subcritical and supercritical primary Hopf bifurcations arise in functionally graded dielectric elastomers.

In conclusion, our study reveals that functional grading is an effective and promising method to broaden the range of controls on the static and dynamic behavior of dielectric elastomers subjected to prestress and time-varying voltages.

## Acknowledgments

GZ gratefully acknowledges the support of GNFM, the Italian Group of Mathematical Physics. MD's work is supported by the Seagull Program of Zhejiang Province, the National Natural Science Foundation of China (No. 11872329), the Natural Science Foundation of Zhejiang Province (No. LD21A020001), and the 111 Project (No. B21034).

---

## References

Alam, Z., & Sharma, A. K. (2022). Functionally Graded Soft Dielectric Elastomer Phononic Crystals: Finite Deformation, Electro-Elastic Longitudinal Waves, and Band Gaps Tunability via Electro-Mechanical Loading. *International Journal of Applied Mechanics*, *14*(06), 2250050. https://doi.org/10.1142/S1758825122500508

Alibakhshi, A., Chen, W., & Destrade, M. (2022). Nonlinear Vibration and Stability of a Dielectric Elastomer Balloon Based on a Strain-Stiffening Model. *Journal of Elasticity*. https://doi.org/10.1007/s10659-022-09893-5

Alibakhshi, A., Dastjerdi, S., Akgöz, B., & Civalek, Ö. (2022). Parametric vibration of a dielectric elastomer microbeam resonator based on a hyperelastic cosserat continuum model. *Composite Structures*, *287*, 115386. https://doi.org/10.1016/j.compstruct.2022.115386

Alibakhshi, A., Dastjerdi, S., Fantuzzi, N., & Rahmanian, S. (2022). Nonlinear free and forced vibrations of a fiber-reinforced dielectric elastomer-based microbeam. *International Journal of Non-Linear Mechanics*, *144*, 104092. https://doi.org/10.1016/j.ijnonlinmec.2022.104092

Alibakhshi, A., Dastjerdi, S., Malikan, M., & Eremeyev, V. A. (2022). Nonlinear free and forced vibrations of a dielectric elastomer-based microcantilever for atomic force microscopy. *Continuum Mechanics and Thermodynamics*. https://doi.org/10.1007/s00161-022-01098-4

Alibakhshi, A., Imam, A., & Haghighi, S. E. (2021). Effect of the second invariant of the Cauchy–Green deformation tensor on the local dynamics of dielectric elastomers. *International Journal of Non-Linear Mechanics*, *137*, 103807. https://doi.org/10.1016/j.ijnonlinmec.2021.103807

Amabili, M. (2008). *Nonlinear Vibrations and Stability of Shells and Plates* (1st ed.). Cambridge University Press. https://doi.org/10.1017/CBO9780511619694

Amabili, M. (2018). *Nonlinear Mechanics of Shells and Plates in Composite, Soft and Biological Materials* (1st ed.). Cambridge University Press. https://doi.org/10.1017/9781316422892

Anssari-Benam, A., & Bucchi, A. (2021). A generalised neo-Hookean strain energy function for application to the finite deformation of elastomers. *International Journal of Non-Linear Mechanics*, *128*, 103626. https://doi.org/10.1016/j.ijnonlinmec.2020.103626

Bazaev, K., & Cohen, N. (2022). Electrically-induced twist in geometrically incompatible dielectric elastomer tubes. *International Journal of Solids and Structures*, *250*, 111707. https://doi.org/10.1016/j.ijsolstr.2022.111707

Behera, S. K., Kumar, D., & Sarangi, S. (2021). Modeling of electro–viscoelastic dielectric elastomer: A continuum mechanics approach. *European Journal of Mechanics - A/Solids*, *90*, 104369. https://doi.org/10.1016/j.euromechsol.2021.104369

Bortot, E., & Shmuel, G. (2018). Prismatic bifurcations of soft dielectric tubes. *International Journal of Engineering Science*, *124*, 104–114. https://doi.org/10.1016/j.ijengsci.2017.11.002

Chen, L., & Yang, S. (2021). Enhancing the Electromechanical Coupling in Soft Energy Harvesters by Using Graded Dielectric Elastomers. *Micromachines*, *12*(10), 1187. https://doi.org/10.3390/mi12101187

Colonnelli S., Saccomandi G., and Zurlo G., Damage induced dissipation in electroactive polymer harvesters, Applied Physics Letters 105, 163904 (2014); doi: 10.1063/1.4900485

Colonnelli, S., Saccomandi, G., & Zurlo, G. (2015). The role of material behavior in the performances of electroactive polymer energy harvesters. *Journal of Polymer Science Part B: Polymer Physics*, *53*(18), 1303–1314. https://doi.org/10.1002/polb.23761

Conroy Broderick, H., Righi, M., Destrade, M., & Ogden, R. W. (2020). Stability analysis of charge-controlled soft dielectric plates. *International Journal of Engineering Science*, *151*, 103280. https://doi.org/10.1016/j.ijengsci.2020.103280

Cooley, C. G., & Lowe, R. L. (2022). In-plane nonlinear vibration of circular dielectric elastomer membranes with extreme stretchability. *European Journal of Mechanics - A/Solids*, *96*, 104660. https://doi.org/10.1016/j.euromechsol.2022.104660

De Tommasi, D., Puglisi, G., & Zurlo, G. (2014). Hysteresis in electroactive polymers. *European Journal of Mechanics - A/Solids*, *48*, 16–22. https://doi.org/10.1016/j.euromechsol.2014.05.011

Dorfmann, L., & Ogden, R. W. (2014). Instabilities of an electroelastic plate. *International Journal of Engineering Science*, *77*, 79–101. https://doi.org/10.1016/j.ijengsci.2013.12.007

Feng, C., Jiang, L., & Lau, W. M. (2011). Dynamic characteristics of a dielectric elastomer-based microbeam resonator with small vibration amplitude. *Journal of Micromechanics and Microengineering*, *21*(9), 095002. https://doi.org/10.1088/0960-1317/21/9/095002

Gent, A. N. (1996). A New Constitutive Relation for Rubber. *Rubber Chemistry and Technology*, *69*(1), 59–61. https://doi.org/10.5254/1.3538357

Ghayesh, M. H., & Farokhi, H. (2015). Chaotic motion of a parametrically excited microbeam. *International Journal of Engineering Science*, *96*, 34–45. https://doi.org/10.1016/j.ijengsci.2015.07.004

Ghosh, A., & Basu, S. (2021). Soft dielectric elastomer tubes in an electric field. *Journal of the Mechanics and Physics of Solids*, *150*, 104371. https://doi.org/10.1016/j.jmps.2021.104371

Guo, Y., Liu, L., Liu, Y., & Leng, J. (2021). Review of Dielectric Elastomer Actuators and Their Applications in Soft Robots. *Advanced Intelligent Systems*, *3*(10), 2000282. https://doi.org/10.1002/aisy.202000282

Gupta, R., & Harursampath, D. (2015). Dielectric elastomers: Asymptotically-correct three-dimensional displacement field. *International Journal of Engineering Science*, *87*, 1–12. https://doi.org/10.1016/j.ijengsci.2014.10.006

Horgan, C. O. (2021). A note on a class of generalized neo-Hookean models for isotropic incompressible hyperelastic materials. *International Journal of Non-Linear Mechanics*, *129*, 103665. https://doi.org/10.1016/j.ijnonlinmec.2020.103665

Jiang, L., Wang, Y., Wang, X., Ning, F., Wen, S., Zhou, Y., Chen, S., Betts, A., Jerrams, S., & Zhou, F.-L. (2021). Electrohydrodynamic printing of a dielectric elastomer actuator and its application in tunable lenses. *Composites Part A: Applied Science and Manufacturing*, *147*, 106461. https://doi.org/10.1016/j.compositesa.2021.106461

Khurana, A., Joglekar, M. M., & Zurlo, G. (2022). Electromechanical stability of wrinkled dielectric elastomers. *International Journal of Solids and Structures*, *246–247*, 111613. https://doi.org/10.1016/j.ijsolstr.2022.111613

Khurana, A., Kumar, A., Raut, S. K., Sharma, A. K., & Joglekar, M. M. (2021). Effect of viscoelasticity on the nonlinear dynamic behavior of dielectric elastomer minimum energy structures. *International Journal of Solids and Structures*, *208–209*, 141–153. https://doi.org/10.1016/j.ijsolstr.2020.10.022

Kim, K. J., & Tadokoro, S. (Eds.). (2007). *Electroactive Polymers for Robotic Applications*. Springer London. https://doi.org/10.1007/978-1-84628-372-7

Liu, L., Liu, Y., & Leng, J. (2013). Theory progress and applications of dielectric elastomers. *International Journal of Smart and Nano Materials*, *4*(3), 199–209. https://doi.org/10.1080/19475411.2013.846281

Lu, T., Ma, C., & Wang, T. (2020). Mechanics of dielectric elastomer structures: A review. *Extreme Mechanics Letters*, *38*, 100752. https://doi.org/10.1016/j.eml.2020.100752

Lv, X., Liu, L., Liu, Y., & Leng, J. (2018). Dynamic performance of dielectric elastomer balloon incorporating stiffening and damping effect. *Smart Materials and Structures*, *27*(10), 105036. https://doi.org/10.1088/1361-665X/aab9db

Mangan, R., & Destrade, M. (2015). Gent models for the inflation of spherical balloons. *International Journal of Non-Linear Mechanics*, *68*, 52–58. https://doi.org/10.1016/j.ijnonlinmec.2014.05.016

Meng, H., & Hu, J. (2010). A Brief Review of Stimulus-active Polymers Responsive to Thermal, Light, Magnetic, Electric, and Water/Solvent Stimuli. *Journal of Intelligent Material Systems and Structures*, *21*(9), 859–885. https://doi.org/10.1177/1045389X10369718

Norris, N.A. Comment on "Method to analyze electromechanical stability of dielectric elastomers". *Appl. Phys. Lett*. https://doi.org/10.1063/1.2833688

Pascon, J. P. (2018). Large deformation analysis of functionally graded visco-hyperelastic materials. *Computers & Structures*, *206*, 90–108. https://doi.org/10.1016/j.compstruc.2018.06.001

Ren, J., & Guo, S. (2021). Thermo-electro-mechanical dynamical response and the failure of an electro-active polymer cylindrical shell. *Applications in Engineering Science*, *5*, 100031. https://doi.org/10.1016/j.apples.2020.100031

Sharma, A. K., Arora, N., & Joglekar, M. M. (2018). DC dynamic pull-in instability of a dielectric elastomer balloon: An energy-based approach. *Proceedings of the Royal Society A: Mathematical, Physical and Engineering Sciences*, *474*(2211), 20170900. https://doi.org/10.1098/rspa.2017.0900

Sharma, A. K., Bajpayee, S., Joglekar, D. M., & Joglekar, M. M. (2017). Dynamic instability of dielectric elastomer actuators subjected to unequal biaxial prestress. *Smart Materials and Structures*, *26*(11), 115019. https://doi.org/10.1088/1361-665X/aa8923

Sheng, J., Chen, H., Li, B., & Wang, Y. (2014). Nonlinear dynamic characteristics of a dielectric elastomer membrane undergoing in-plane deformation. *Smart Materials and Structures*, *23*(4), 045010. https://doi.org/10.1088/0964-1726/23/4/045010

Su, Y, Broderick, H C, Chen, W, Destrade, M. Wrinkles in soft dielectric plates. *Journal of the Mechanics and Physics of Solids*. 119, 298--318. https://doi.org/10.1016/j.jmps.2018.07.001

Su, Y. (2020). Voltage-controlled instability transitions and competitions in a finitely deformed dielectric elastomer tube. *International Journal of Engineering Science*, *157*, 103380. https://doi.org/10.1016/j.ijengsci.2020.103380

Su, Y., Ogden, R. W., & Destrade, M. (2021). Bending control and stability of functionally graded dielectric elastomers. *Extreme Mechanics Letters*, *43*, 101162. https://doi.org/10.1016/j.eml.2020.101162

Suo, Z. (2010). Theory of dielectric elastomers. *Acta Mechanica Solida Sinica*, *23*(6), 549–578. https://doi.org/10.1016/S0894-9166(11)60004-9

DeTommasi, D., Puglisi, G., & Zurlo, G. (2014). Failure modes in electroactive polymer thin films with elastic electrodes. *Journal of Physics D: Applied Physics*, *47*(6), 065502. https://doi.org/10.1088/0022-3727/47/6/065502

Wang, F., Lu, T., & Wang, T. J. (2016). Nonlinear vibration of dielectric elastomer incorporating strain stiffening. *International Journal of Solids and Structures*, *87*, 70–80. https://doi.org/10.1016/j.ijsolstr.2016.02.030

Wu, B., Destrade, M., & Chen, W. (2020). Nonlinear response and axisymmetric wave propagation in functionally graded soft electro-active tubes. *International Journal of Mechanical Sciences*, *187*, 106006. https://doi.org/10.1016/j.ijmecsci.2020.106006

Xia, G., Su, Y., & Chen, W. (2021). Instability of compressible soft electroactive plates. *International Journal of Engineering Science*, *162*, 103474. https://doi.org/10.1016/j.ijengsci.2021.103474

Xu, B.-X., Mueller, R., Theis, A., Klassen, M., & Gross, D. (2012). Dynamic analysis of dielectric elastomer actuators. *Applied Physics Letters*, *100*(11), 112903. https://doi.org/10.1063/1.3694267

Yin, Y., Zhao, D., Liu, J., & Xu, Z. (2022). Nonlinear dynamic analysis of dielectric elastomer membrane with electrostriction. *Applied Mathematics and Mechanics*, *43*(6), 793–812. https://doi.org/10.1007/s10483-022-2853-9

Yong, H., He, X., & Zhou, Y. (2011). Dynamics of a thick-walled dielectric elastomer spherical shell. *International Journal of Engineering Science*, *49*(8), 792–800. https://doi.org/10.1016/j.ijengsci.2011.03.006

Zhang, J., & Chen, H. (2020). Voltage-induced beating vibration of a dielectric elastomer membrane. *Nonlinear Dynamics*, *100*(3), 2225–2239. https://doi.org/10.1007/s11071-020-05678-4

Zhang, J., Chen, H., & Li, D. (2016). Method to Control Dynamic Snap-Through Instability of Dielectric Elastomers. *Physical Review Applied*, *6*(6), 064012. https://doi.org/10.1103/PhysRevApplied.6.064012

Zhang, J., Chen, H., & Li, D. (2018). Pinnacle elimination and stability analyses in nonlinear oscillation of soft dielectric elastomer slide actuators. *Nonlinear Dynamics*, *94*(3), 1907–1920. https://doi.org/10.1007/s11071-018-4464-y

Zhao, X., Koh, S. J. A., & Suo, Z. (2011). Nonequilibrium thermodynamics of dielectric elastomers. *International Journal of Applied Mechanics*, *03*(02), 203–217. https://doi.org/10.1142/S1758825111000944

Zhou, W., Chen, Y., & Su, Y. (2020). Bifurcation of a finitely deformed functionally graded dielectric elastomeric tube. *International Journal of Non-Linear Mechanics*, *127*, 103593. https://doi.org/10.1016/j.ijnonlinmec.2020.103593

Zhu, J., Cai, S., & Suo, Z. (2010). Nonlinear oscillation of a dielectric elastomer balloon. *Polymer International*, *59*(3), 378–383. https://doi.org/10.1002/pi.2767

Zou, H.-L., Deng, Z.-C., & Zhou, H. (2022). Revisited chaotic vibrations in dielectric elastomer systems with stiffening. *Nonlinear Dynamics*, *110*(1), 55–67. https://doi.org/10.1007/s11071-022-07617-x

Zurlo, G., Destrade, M., & Lu, T. (2018). Fine tuning the electro-mechanical response of dielectric elastomers. *Applied Physics Letters*, *113*(16), 162902. https://doi.org/10.1063/1.5053643